# A PROPOSAL TO CLASSIFY LATINAMERICAN SCIENTIFIC JOURNALS USING CITATION INDICATORS: CASE STUDY IN COLOMBIA

**ROMERO-TORRES, M**

Unidad de Ecología y Sistemática (UNESIS), Departamento de Biología, Pontificia Universidad Javeriana, Bogotá, Colombia. Carrera 7 No 43-82. Teléfono: +57 (1) 3208320 Ext. 4082. Research ID: A-9086-2009. E-mail: mauricio_romero@javeriana.edu.co

**TEJADA, MA**

Coordinadora de Publicaciones Seriadas, Editorial Pontificia Universidad Javeriana, Pontificia Universidad Javeriana, Bogotá, Colombia. Carrera 7 No 43-82. Teléfono: +57 (1) 3208320 Ext. 4755. E-mail: maria.tejada@javeriana.edu.co

**ACOSTA, A**

Unidad de Ecología y Sistemática (UNESIS), Departamento de Biología, Pontificia Universidad Javeriana, Bogotá, Colombia. Carrera 7 No 43-82. Teléfono: +57 (1) 3208320 Ext. 4082. E-mail: laacosta@javeriana.edu.co

**ABSTRACT**

Colombian scientific journals are poorly represented in international digital libraries; however, through Google Scholar (GS) it is possible to determine their use by the community. Between the years of 2003 and 2007 a classification of 185 Colombian journals indexed in the Colombian National Bibliographical Index (IBNP) was performed using the information provided by GS, basing categorization on size indicators, indexation and citation. The indicators were analyzed by grouping the journals in two general areas: sciences and social sciences. In each area, the indicators provided by the digital libraries Scopus, Redalyc and Scielo were compared. Additionally, the indicators provided by IBNP journals categories (A1, A2, B and C) were also compared. The sciences and social sciences had a similar pattern in their indicators. The existence of positive correlations was established between some indicators and they predicted that the number of citations per journal in GS and the h index depends on its visibility in GS and Scopus. We put forward that the current IBNP categories (A1, A2, B or C) faintly reflect the use of journals by the community and we propose a classification based on the h index as an infometric indicator, which reflects not only its visibility in Google Scholar, but also its inclusion in certain international digital libraries, particularly Scopus. Our results may be applied to the creation of public policies regarding science and technology in Colombia and in developing countries.

# PROPUESTA PARA CLASIFICAR REVISTAS CIENTIFICAS EN LATINOAMERICA MEDIANTE INDICADORES DE CITACIÓN: ESTUDIO DE CASO EN COLOMBIA

## RESUMEN

Las revistas científicas colombianas son escasamente representadas en librerías digitales internacionales; sin embargo, a través de Google Académico (GA) es posible determinar el uso que la comunidad hace de estas. Utilizando GA se realizó entre los años 2003 y 2007 una clasificación de 185 revistas colombianas indexadas del Índice Bibliográfico Nacional-Publindex de Colombia (IBNP), a través de indicadores de tamaño, indexación y citación. Los indicadores se analizaron agrupando las revistas en dos áreas generales: ciencias y ciencias sociales. En cada área, se compararon indicadores entre las librerías digitales Scopus, Redalyc y Scielo. Así mismo, se compararon indicadores entre las categorías de revistas IBNP (A1, A2, B y C). Ciencias y ciencias sociales tuvieron un patrón similar en sus indicadores. Se estableció la existencia de correlaciones positivas entre algunos indicadores y se predijo que la cantidad de citas por revista en GA y el índice $h$, dependen de su visibilidad en GA y Scopus. Se sugiere que las actuales categorías del IBNP (A1, A2, B o C) reflejan débilmente el uso de las revistas por parte de la comunidad y proponemos una clasificación basada en el índice $h$ como indicador infométrico, el cual refleja tanto su visibilidad en Google Académico, como la pertenencia a ciertas librerías digitales mundiales, particularmente Scopus. Nuestros resultados pueden ser aplicados en la formulación de políticas públicas de ciencia y tecnología de Colombia y países en vía de desarrollo.

**Palabras clave**

Google Académico, Índice $h$, Publindex, Colombia, políticas, infometría

## INTRODUCCION

Colombia, recientemente modificó sus políticas públicas relacionadas con el sistema de ciencia y tecnología (Ley 1289 de 2009), invocando la necesidad de establecer mecanismos de evaluación de los resultados de investigación de instituciones públicas y privadas. Este cambio de políticas impulsó en las instituciones la comparación de su productividad y calidad en el ámbito local, regional y global, por medio de diversos indicadores como los presentados en las clasificaciones mundiales de universidades (e.g. clasificación de Shanghai), clasificación iberoamericana (e.g. Ranking Iberoamericano SIR 2010) o colombiana a través del Atlas Colombiano de la Ciencia (http://www.scimagojr.com/atlascolombia), entre otros.

Ante los resultados de estas clasificaciones, deletéreos para la mayoría de las instituciones colombianas y latinoamericanas debido a que muy pocas aparecen clasificadas, se han planteado estrategias institucionales que buscan revertir la tendencia de estos indicadores. Esto puede interpretarse como el ingreso formal de la ciencia colombiana en todas sus escalas, desde investigadores hasta instituciones y revistas académicas, al sistema de evaluación por indicadores de citación.

Aunque el uso de estos indicadores es controversial (Alfonso et al. 2005; Camps 2008; Todd & Ladle 2008), a través de medidas como la cantidad de artículos publicados, cantidad de citas, el factor de impacto (Todd 2009) y recientemente los índices Eigenfactor, Scimago Journal Rank, SNIP e índice *h*, entre otros (Torres-Salinas & Jimenez-Contreras 2010), se mide y compara el desempeño en investigación de individuos, grupos e instituciones (Van Leeuwen et al. 2003).

La citación es parte formal del proceso científico (Egghe & Rousseau 1990). Es el reconocimiento positivo o negativo que hace la comunidad científica a una pieza de conocimiento (Merton 1968). Bajo el supuesto que un número alto de citaciones es sinónimo de calidad científica (Egghe & Rousseau 1990; Van Leeuwen *et al.* 2003), artículos, autores y revistas altamente citados ejercen una mayor influencia intelectual sobre un área particular (Merton 1968) y se asume que estos presentan aportes, avances o una síntesis significativa para el conocimiento y por lo tanto, se convierten en referentes teóricos o experimentales para una disciplina (Bensman 2007; Merton 1968; Merton & Sztompka 1996; Van Leeuwen *et al*. 2003).

El uso que la comunidad hace de los artículos y revistas académicas está relacionado con su visibilidad, es decir, el éxito que tiene una publicación al emplear los mecanismos y medios digitales de divulgación en la red mundial de información, para lograr que un usuario la encuentre, acceda, evalue, emplee, retroalimente y finalmente, cite. Esto significa que la visibilidad de una publicación es influida por diversos factores como: (i) la presencia en librerías digitales o librería digital globales de indexación (Hull *et al.* 2008); (ii) capacidad de ser consultada y reconocida por la comunidad científica mundial (Figueira *et al.* 2003); (iii) la composición del consejo editorial que respalda el proceso de revisión por pares (Ren & Rousseau 2002); (iv) su patrón de citación que determina su posicionamiento en los buscadores (Chen et al. 2007; Todd & Ladle 2008) y posterior citación (Katz 1999; Leydesdorff & Bensman 2006); la pertenencia de autores y consejo editorial a sociedades científicas (Leydesdorff & Bensman 2006) o pequeños mundos (Newman 2003) y (v) la cantidad relativa de producción de artículos en un área a nivel mundial (Huberman 2001; Ingwersen 2000), entre otros.

Es intuitivo encontrar que la visibilidad, citación o influencia intelectual es dominada por Estados Unidos, Inglaterra, Canadá y otros países desarrollados (Adams 1998), porqué controlan los factores anteriormente expuestos y además, porqué en estos países las publicaciones académicas son en sí mismas una industria (Bergstrom 2004; Bergstrom & Bergstrom 2006).

En países en desarrollo, la mayoría de las publicaciones académicas y personal científico están a la periferia de estos procesos de comunicación, sus servicios y beneficios (Gevers 2009). Esto permite inducir (respecto a los países desarrollados) que el modelo de comunicación de las revistas académicas de Latinoamérica o de países en desarrollo, se caracteriza por:

- El bajo reconocimiento internacional de sus avances de investigación (Gevers 2009) consecuencia de la calidad (creatividad, originalidad y contribución en un área) de sus artículos (Arunachalam & Manorama 1989; Zhou & Leydesdorff 2007).
- Barrera lingüística de los autores (Bertrand & Hunter 1998; Zhou & Leydesdorff 2007)
- Baja disponibilidad de acceso en línea (Zhou & Leydesdorff 2007) y pocas revistas están indexadas en los principales sistemas de indexación (e.g. Web of Knowledge o Scopus) (Gorbea-Portal & Suárez-Balseiro 2007).
- Los estándares de calidad editorial son altamente variables (Zhou & Leydesdorff 2007). Hay una proliferación de revistas de corta vida, baja calidad editorial y científica, con pocos canales de distribución (Gevers 2009).
- Existen relaciones endogámicas entre países y revistas como reflejo de lo que ocurre en el resto del sistema de comunicación científica que integran estas revistas (Gorbea-Portal & Suárez-Balseiro 2007).

- La mayor proporción de la producción nacional se publica en revistas domésticas, por lo cual, una considerable cantidad de publicaciones no están incluidas en índices de citación (Bertrand & Hunter 1998) llevando a que los autores sean invisibles al resto del mundo (Figueira *et al.* 2003).
- Las revistas no son una industria y hacen énfasis en sostener la actividad de investigación local (Gevers 2009).
- La visibilidad internacional depende de la colaboracion con pares extranjeros (Gevers 2009; Jimenez-Contreras *et al.* 2010) que publican sus trabajos en revistas generalmente no latinoamericanas.

Las revistas Colombianas no son ajenas a las tendencias de cambio, indexación, globalización y comparación, como lo demuestran algunos editoriales nacionales (Laverde & Clara 2009; Lopez-Lopez 2009; Lopez-Lopez 2010) y al parecer Colombia se encuentra en una etapa de transición desde un sistema de divulgación e incentivos que hace énfasis en publicaciones domésticas (cerrado y endogámico), hacia uno visible internacionalmente que se soporta en redes de conocimiento y es medido por indicadores de citación.

Actualmente Colombia cuenta con la Base Bibliográfica Nacional Publindex (BBNP) que regula la indexación, clasificación y homologación de más de trescientas revistas académicas de instituciones y asociaciones colombianas. El sistema tiene cuatro categorías de clasificación (A1, A2, B y C); sin embargo, hasta el 2007 no ha empleado sensu stricto indicadores de citación para construir la clasificación de revistas, debido a la inclusión de pocas revistas en librerías digitales internacionales que permiten el cálculo de indicadores citación (e.g. Web of Knowledge o Scopus).

No obstante esta situación, actualmente es posible estimar la citación de revistas académicas a través del servicio de Google Académico (Harzing & Wal 2009; Leydesdorff 2009; Norris & Oppenheim 2010), el cual realiza búsquedas en un rango amplio de editoriales académicas, sociedades profesionales, repositorios, universidades y otras organizaciones escolares (Jacso 2005). Con base en lo anterior, para las revistas colombianas indexadas en IBNP se plantearon los siguientes objetivos: entre los años 2003 a 2007 (i) estimar y comparar indicadores de tamaño, indexación y citación, (ii) establecer relaciones entre estos indicadores, y (iii) realizar una clasificación de las revistas colombianas.

**MÉTODOS**

**Estimación de indicadores de tamaño, indexación y citación**

Durante el mes de agosto de 2008 se consultó el Índice Bibliográfico Nacional Publindex (IBNP) 2008-I (http://scienti.colciencias.gov.co:8084/publindex) y su actualización. Se seleccionaron 209 revistas, de las cuales nueve pertenecían a la categoría A1, cuarenta y dos a la categoría A2 y cincuenta a la categoría B. Para la categoría C se tomó una muestra representativa de 108 revistas ($\alpha$ = 0.05, $\beta$ = 0.9). En el portal de búsqueda del IBNP, entre los años 2003 a 2007 se realizó un conteo del tamaño o cantidad de artículos publicados por cada revista (cantidad de artículos por revista indexados en IBNP - ARIIBNP). Se definió artículo en su sentido más amplio y se incluyeron todos los ítems disponibles como artículos originales, artículos de revisión, editoriales, cartas al editor, traducciones y revisiones de libro. La sumatoria del total de artículos en IBNP para el quinquenio se consideró como la producción total de artículos publicados.

Bajo el supuesto que las revistas no son homogéneas entre sí respecto a sus áreas de conocimiento y que existen pocas revistas para realizar una clasificación en áreas específicas, las revistas se agruparon en dos áreas de conocimiento general: Ciencias y Ciencias Sociales.

Para los indicadores de indexación, se consultó la presencia o ausencia de cada revista en las librerías digitales internacionales por suscripción: Thomson Reuters Web of Knowledge (WoK) (http://apps.isiknowledge.com) y Scopus (http://www.scopus.com); en las librerías digitales regionales de libre acceso Redalyc (http://redalyc.uaemex.mx) y Scielo (http://www.scielo.org) y en la librería digital Google Académico (http://scholar.google.com) a través del programa Publish or Perish (Harzing & Wal 2009).

Con base en los datos de presencia o ausencia, se estimó el puntaje de indexación de librería digital ($PI_{LD}$), bajo el supuesto que las revistas indexadas en el WoK o Scopus tenían un mayor puntaje. Se asignaron puntajes arbitrariamente con incrementos siguiendo una función de potencia (1, 10 y 100). Se asignó un puntaje de 100 a revistas indexadas en WoK o Scopus, un puntaje de diez a revistas indexadas en Redalyc y Scielo, y un puntaje de uno a revistas indexadas en Google Académico (puntaje mínimo posible = 0 y máximo posible = 221). Para estimar el puntaje de indexación IBNP ($PI_{IBNP}$), se tuvo el supuesto que a mayor categoría IBNP, mayor puntaje. Utilizando el listado del IBNP, a cada revista se asignó arbitrariamente un puntaje de 4, 3, 2 y 1 de acuerdo a su categoría A1, A2, B y C, respectivamente.

Para determinar los indicadores de citación, en primer lugar se consultó el Journal Citation Report de Thomson Reuters Web of Knowledge del año 2009, seleccionando las revistas colombianas indexadas en el Science Citation Index y en el Social Science Citation Index. Para cada revista se determinó la cantidad de artículos indexados por revista (*AIR*) y la cantidad de citas por revista (*CRR*). En la librería digital Scopus, se realizó para cada revista un análisis similar a través de la opción "registro de citación" (en inglés: Citation Tracker) teniendo en cuenta la auto citación. Se tuvo en cuenta la cantidad artículos indexados por revista, la cantidad de citas por revista y el índice *h*.

Posteriormente, se consultó Google Académico utilizando el programa Publish or Perish (GA-PoP) (Harzing & Wal 2009). Mediante la opción "Análisis de Impacto de Revistas" (en inglés: Journal Impact Analysis) se tuvieron en cuenta todas las áreas de conocimiento y el rango temporal 2003 a 2007. La búsqueda utilizó el título exacto de la revista (incluyendo tildes). Para cada revista se consultó la cantidad de artículos indexados por revista ($AIR_{GA}$), la cantidad de citas por revista ($AIR_{GA}$) y el índice *h* ($h_{GA}$). Manualmente eliminaron las referencias duplicadas teniendo como criterios: (i) títulos de revistas similares en la columna "Publicación" (en inglés: Publication); (ii) duplicidad del título en idioma inglés de un ítem ya citado en idioma español y (iii) referencias incompletas con campos vacios. Para los ítems con algún grado de incertidumbre, se consultó la validez de la referencia accediendo a su vínculo en Google Académico y en algunos casos, se consultó directamente la fuente primaria.

En cada revista se estimó la proporción entre la cantidad de artículos indexados en IBNP (producción real), respecto a la que aparece indexada en GA (producción visible), por medio de la relación ($AIR_{GA}$ /$AIR_{IBNP}$). Así mismo, para cada revista indexada en GA, se calculó la citación promedio por artículo ($CA\dot{x}_{GA}$) como la relación: $CRR_{GA}$ / $AIR_{IBNP}$.

Para determinar la existencia de diferencias significativas al comparar indicadores entre librerías digitales y entre categorías de revistas IBNP, se utilizó un análisis de varianza de una vía cuidando cumplir con los supuestos $\varepsilon_i$ independientes $N$ (0, $\sigma^2$), o su equivalente no paramétrico o prueba de comparación múltiple por rangos de Kruskal-Wallis y prueba post-hoc de Tukey. Se excluyeron de la comparación de librerías digitales los valores de WoK (n = 1) Los valores de cantidad de artículos indexados por revista en IBNP ($AIR_{IBNP}$) y cantidad de citas por revista en GA ($CR_{GA}$), se transformaron a $Log_{10}$ para reducir su varianza (a los valores de cero citas se sumó uno para la transformación). En todos los casos se utilizó un nivel de significancia del 95 %.

| Indicadores empleados | Símbolo |
|---|---|
| Cantidad de artículos indexados por revista en IBNP, GA y Scopus | $AIR_{IBNP}$, $AIR_{GA}$, $ARI_{SC}$ |
| Cantidad de artículos promedio indexados por revista en GA y Scopus | $AIR\dot{x}_{IBNP}$, $AIR\dot{x}_{GA}$, $ARI\dot{x}_{SC}$ |
| Cantidad de citas por revista en GA y Scopus | $CR_{GA}$, $CR_{SC}$ |
| Citas promedio por artículo en GA | $CA\dot{x}_{GA}$ |
| Citas promedio por artículo en el área | $CA\dot{x}_{AREA}$ |
| Índice h en *Google Académico* | $h_{GA}$ |
| Índice h en *Scopus* | $h_{SC}$ |
| Índice de citación promedio normalizado por artículo | $CPN$ |
| Puntaje de Indexación de librería digital | $PI_{LD}$ |
| Puntaje de Indexación IBNP | $PI_{IBNP}$ |

**Relaciones de asociación y dependencia entre variables**

Para cada área general, se realizaron pruebas de correlación entre las variables de tamaño (cantidad de artículos indexados por revista en IBNP), indexación (puntajes de indexación IBNP y de librería digital) y citación (citas por revista en GA, citas promedio por artículo en GA e índice h en GA).

Se verificó el cumplimiento de los supuestos paramétricos ($\varepsilon_i$ independientes $N(0, \sigma^2)$) y en caso contrario, se realizó una prueba de correlación por rangos de Spearman ($r_s$). Se utilizó un nivel de significancia del 95 %. Los valores de $AIR_{IBNP}$, $AIR_{GA}$ y $CR_{GA}$, se transformaron a $Log_{10}$ para reducir su varianza (Leydesdorff & Bensman 2006).

Ya que las variables indicadoras de citación no son independientes entre sí, se determinó su grado de dependencia lineal para determinar cuál de estas explica mejor la variabilidad de los datos. Se realizó un análisis de factores (Leydesdorff 2006; Leydesdorff 2007) entre los índices $h_{GA}$, $Log_{10}$ $CR_G$ y $CPA_{GA}$ utilizando como método de extracción análisis de componentes principales sin rotación. Se consideró un componente significante, aquel con un valor propio ($\lambda$) mayor a 1, el valor de comunalidad o cantidad de varianza aportada por cada variable al factor mayor que 0.80 y un factor de carga (contribución única de cada variable al factor) mayor a 0.7 (Hair 1995; Hardle & Hlávka 2007).

Con el fin de establecer la relación de dependencia lineal entre el cambio promedio de los indicadores de citación y las variables de indexación y tamaño, se llevó a cabo una regresión lineal múltiple. Las regresiones emplearon suma de cuadrados Tipo I (secuenciales), se cumplieron los supuestos de una regresión ($\varepsilon_i$ independientes $N$ (0, $\sigma^2$)) y de colinealidad (Chatterjee *et al.* 2000), utilizando en todos los casos un nivel de confianza del 95 %.

**Clasificación de las revistas colombianas**

Las revistas de las áreas generales de ciencias y ciencias sociales, se organizaron en orden descendente con base en su valor de índice $h_{GA}$ y posteriormente se dividieron en cuartiles. Se presentarán exclusivamente las revistas de los primeros dos cuartiles (valor del índice $h_{GA}$ que supera el 50 % de las observaciones o mediana).

Así mismo, en cada área se calculó para cada revista el índice de citación promedio normalizado (*CPN*), como la relación entre las citas promedio por artículo en GA y el promedio de citas por artículo del área (*CPA$_{AREA}$*)), equivalentes los índice a *CPP y FCS* propuestos y corregidos por Moed (1995) y Opthof (2010) respectivamente. El numerador está también normalizado para evitar un efecto del tamaño de las revistas sobre la citación (Opthof & Leydesdorff 2010; Radicchi et al. 2008). La relación indica citas observadas/citas esperadas, de tal forma, si la relación es > 1.0 (e.g. 3.0), significa que la tasa de citación promedio de un artículo de esa revista, es tres veces mayor al promedio nacional en el área.

**RESULTADOS**

**Análisis de citación de las revistas colombianas**

Para 209 revistas pertenecientes al IBNP 2008-I, se realizó un recuento de 25490 artículos. Se realizaron conteos para 111 revistas del área general de ciencias y 98 revistas de ciencias sociales. Fue posible estimar índices de citación para 185 revistas (88.5 %) en GA a través del software Publish or Perish. Del total de artículos considerados, GA fue la librería digital que indexó una mayor proporción de ítems (9035), equivalente al 35.34 % (Tablas 1 y 2). Las restantes librería digital indexaron porcentajes menores.

**Indicadores área general ciencias**

Para las 111 revistas del área de ciencias, se hizo un recuento de 15730 ítems en el IBNP. En términos del tamaño, GA indexó la mayor cantidad de artículos (35.43 %) respecto a las demás librerías digitales (Tabla 1). Diez revistas no fueron indexadas en GA y nueve revistas no se encontraron indexadas en alguna de las librerías digitales consideradas. El índice $h_{GA}$ de revistas presentó valores entre 0 y 10.

**Tabla 1.** Valores globales de tamaño, indexación y citación de las revistas colombianas para el área general ciencias, indexadas en las librerías digitales WoK, Scopus, Redalyc, Scielo, y Google Académico. [a, b,] Indican grupos homogéneos después de pruebas de comparación múltiple.

| Indicador | WoK | SCOPUS | Redalyc | Scielo | Google Académico |
|---|---|---|---|---|---|
| **Cantidad de revistas indexadas** | 1 | 10 | 19 | 42 | 101 |
| **Cantidad de artículos de IBNP en GA-PoP y porcentaje del total** | 126 (0.08 %) | 1520 (9.66 %) | 1421 (14.21 %) | 3336 (21.21 %) | 5574 (35.43 %) |
| **Citas totales recibidas en GA-PoP** | 69 | 1880 | 1335 | 2271 | 3499 |
| **Citas promedio por revista en GA-PoP (Log $_{10}$) y desviación estándar** | 1.84 | 1.65 [a] (0.92) | 1.23 [a] (0.70) | 1.21 [a] (0.66) | 1.06 [b] (0.64) |
| **Citas promedio por artículo en GA-PoP y desviación estándar** | 0.31 | 0.52 (0.62) | 0.34 (0.47) | 0.27 (0.42) | 0.20 (0.34) |

Al comparar los indicadores de citación entre las librerías digitales, las citas promedio (Log$_{10}$) por revista fueron menores en Google Académico respecto a las demás (F = 3,27; gl = 3; $p < 0,05$), pero no existieron diferencias entre las restantes librerías digitales. Respecto a las citas promedio por artículo, no existieron diferencias entre librerías digitales (H (3) = 5.90; n = 166; $p > 0.05$).

Al comparar los indicadores de tamaño, indexación y citación entre las categorías de revistas IBNP (A1, A2, B y C) para todas las variables consideradas (Tabla 2), las categorías se agruparon en dos grupos homogéneos: A1-A2 y B-C. Las revistas A1-A2 presentaron, en la mayoría de variables consideradas (excepto en la relación promedio GA/IBNP) valores mayores respecto a las revistas B-C (en todos los casos $p < 0.05$).

**Tabla 2.** Valores totales y promedios del tamaño, indexación y citación para las revistas del área general ciencias en IBNP para las categorías A1, A2, B y C. Valores entre paréntesis indican desviación estándar. [a, b, c] Indican grupos homogéneos después de pruebas de comparación múltiple.

| Categoría IBNP | Cantidad de revistas | Tamaño | | | | Indexación | Citación | | |
|---|---|---|---|---|---|---|---|---|---|
| | | $AIR_{IBNP}$ (A) | $AIR_{GA}$ (B) | Relación B/A | $AIR\dot{x}_{IBNP}$ ($Log_{10}$) | $PI_{LD}$ | Citas totales recibidas GA | $CR_{GA}$ ($Log_{10}$) | $CA\dot{x}_{GA}$ |
| A1 | 4 | 1157 | 480 | 0.43 [b] (0.35) | 2.45 [a] (0.13) | 93.50 [a] (48.56) | 492 | 1.75 [a] (0.80) | 0.39 [a] (0.29) |
| A2 | 27 | 4885 | 2333 | 0.47 [a] (0.37) | 2.18 [a] (0.25) | 31.37 [a] (40.04) | 1590 | 1.39 [a] (0.57) | 0.36 [a] (0.51) |
| B | 30 | 3716 | 1076 | 0.28 [b] (0.30) | 1.99 [b] (0.29) | 15.96 [b] (32.67) | 972 | 0.89 [b] (0.61) | 0.12 [b] (0.22) |
| C | 50 | 5612 | 1685 | 0.30 [b] (0.28) | 1.92 [b] (0.32) | 3.62 [b] (4.66) | 498 | 0.79 [b] (0.52) | 0.09 [b] (0.14) |

**Indicadores área general ciencias sociales**

Para 98 revistas del IBNP se realizó el recuento en de 10120 ítems. Respecto al tamaño, Redalyc indexó la mayor cantidad de revistas, pero GA-PoP indexó la mayor cantidad de ítems (Tabla 2); sin embargo, catorce revistas indexadas en Redalyc, no presentaron registros en GA-PoP. Una revista no estuvo indexada en alguna de las librerías digitales consultadas. El índice $h_{GA}$ de revistas presentó valores entre 0 y 8.

La comparación de indicadores de citación entre librerías digitales, mostró que el $Log_{10}$ de las citas promedio por revista y citas promedio por artículo en GA, fueron menores en Redalyc y GA-PoP, que en Scopus y Scielo (F = 4.31; gl = 3; *p* < 0.05) y (F = 4.05; gl = 3; *p* < 0.05), respectivamente.

**Tabla 3.** Valores de citación de las revistas colombianas del área general ciencias sociales, indexadas en las librerías digitales WoK, Scopus, Redalyc, Scielo y Google Académico.

| Indicador | WoK | SCOPUS | Redalyc | Scielo | Google Académico |
|---|---|---|---|---|---|
| **Cantidad de revistas indexadas** | 1 | 6 | 89 | 25 | 84 |
| **Cantidad de artículos de IBNP en GA-PoP y porcentaje del total** | 104 (1.03 %) | 481 (4.75 %) | 3262 (32.23 %) | 1549 (15.31 %) | 3461 (34.20 %) |
| **Citas totales recibidas en GA-PoP** | 220 | 544 | 2084 | 1197 | 2227 |
| **Citas promedio por revista en GA-PoP ($Log_{10}$) y desviación estándar** | 2.34 | 1.83 [a] (0.36) | 1.15 [b] (0.58) | 1.44 [a] (0.55) | 1.13 [b] (0.59) |
| **Citas promedio por artículo en GA-PoP y desviación estándar** | 0.96 | 0.57 [a] (0.26) | 0.25 [b] (0.28) | 0.34 [a] (0.30) | 0.25 [b] (0.27) |

La comparación de los indicadores de tamaño, indexación y citación (Tabla 4), entre las cuatro categorías de revistas IBNP en todas las variables consideradas, presentó que las revistas categoría A1 son diferentes en todas las variables de las revistas categoría C, excepto en la relación promedio B/A. Esta relación B/A indicó la relación de cuantos artículos producidos son visibles en Google Académico. Sin embargo, las diferencias entre categorías A1-A2 y A2-B-C dependen de la variable analizada.

**Tabla 4.** Valores de tamaño, indexación y citación para las revistas del área general ciencias sociales del IBNP en las categorías A1, A2, B y C. Valores entre paréntesis indican desviación estándar. [a, b, c] Indican grupos homogéneos después de pruebas de comparación múltiple.

| Categoría IBNP | Tamaño | | | | | Indexación | Citación | | |
|---|---|---|---|---|---|---|---|---|---|
| | Cantidad de revistas | $AIR_{IBNP}$ (A) | $AIR_{GA}$ (B) | Relación B/A | $AIR\dot{x}_{IBNP}$ ($Log_{10}$) | $PI_{LD}$ | Citas totales recibidas GA | $CR_{GA}$ ($Log_{10}$) | $CA\dot{x}_{GA}$ |
| A1 | 5 | 918 | 432 | 0.47 [a] (0.34) | 2.23 [a] (0.20) | 121.00 [a] (70.71) | 527 | 1.93 [a] (0.31) | 0.59 [a] (0.34) |
| A2 | 15 | 1690 | 980 | 0.57 [a] (0.33) | 2.01 [abc] (0.20) | 31.00 [a] (34.84) | 715 | 1.55 [a] (0.42) | 0.45 [a] (0.35) |
| B | 20 | 1724 | 579 | 0.33 [a] (0.35) | 1.90 [bc] (0.18) | 14.50 [a] (4.89) | 315 | 1.02 [b] (0.45) | 0.18 [ab] (0.21) |
| C | 44 | 5788 | 1470 | 0.37 [a] (0.32) | 1.89 [bc] (0.26) | 9.72 [b] (3.07) | 670 | 0.94 [b] (0.55) | 0.17 [b] (0.19) |

**Relaciones entre indicadores**

En las áreas de ciencias y ciencias sociales se presentaron correlaciones positivas débiles (0.3 < r < 0.6), intermedias (0.6 < r < 0.8) y fuertes (0.8 < r < 0.9) entre indicadores de tamaño, indexación y citación (Tabla 5). Fue evidente la asociación positiva entre el índice $h_{GA}$ con todos los indicadores.

En el área general de ciencias, los indicadores de tamaño se correlacionaron débilmente con los indicadores de indexación, y de forma débil e intermedia con los indicadores de citación. Los indicadores de indexación, en general se asociaron débilmente con los restantes indicadores. Por último, los indicadores de citación presentaron correlaciones débiles, intermedias y fuertes con los indicadores de tamaño e indexación.

Los indicadores de citación no fueron linealmente independientes (se excluyó el *índice h* de Scopus) y se encuentran en el mismo espacio vectorial. El análisis de factores mostró que un solo factor explicó la combinación lineal de las tres variables. El índice $h_{GA}$ explicó la mayor cantidad de varianza del factor (89.28 %) y así mismo, su factor de carga o contribución única de cada variable al factor, fue también la más alta (0.94).

**Tabla 5.** Valores del análisis de correlación de Spearman de los indicadores de tamaño, indexación y citación para las revistas del área general ciencias. Ns significa no significativo ($p > 0.05$).

| Indicador | $AIR_{IBNP}$ (Log$_{10}$) | $AIR_{GA}$ (Log$_{10}$) | $PI_{IBNP}$ | $PI_{LD}$ | $CR_{GA}$ (Log$_{10}$) | $CA\dot{x}_{GA}$ | $h_{GA}$ | $h_{SC}$ |
|---|---|---|---|---|---|---|---|---|
| $AIR_{IBNP}$ (Log$_{10}$) | 1.00 | 0.47 | 0.42 | 0.41 | 0.46 | Ns | 0.43 | Ns |
| $AIR_{GA}$ (Log$_{10}$) | 0.47 | 1.00 | 0.45 | 0.44 | 0.81 | 0.70 | 0.73 | 0.77 |
| $PI_{IBNP}$ | 0.42 | 0.45 | 1.00 | 0.63 | 0.42 | 0.35 | 0.41 | Ns |
| $PI_{LD}$ | 0.41 | 0.44 | 0.63 | 1.00 | 0.36 | 0.45 | 0.41 | Ns |
| $CR_{GA}$ (Log$_{10}$) | 0.46 | 0.81 | 0.42 | 0.36 | 1.00 | 0.89 | 0.93 | 0.74 |
| $CA\dot{x}_{GA}$ | Ns | 0.70 | 0.35 | 0.45 | 0.89 | 1.00 | 0.84 | Ns |
| $h_{GA}$ | 0.43 | 0.73 | 0.41 | 0.41 | 0.93 | 0.84 | 1.00 | 0.68 |
| $h_{SC}$ | Ns | 0.77 | Ns | Ns | 0.74 | Ns | 0.68 | 1.00 |

En el área de ciencias sociales, se encontraron asociaciones positivas débiles, intermedias y fuertes entre los indicadores de tamaño, indexación y citación (Tabla 6). Es de destacar que solo existió una correlación intermedia entre el puntaje de indexación de IBNP y el puntaje de indexación de librería digital, así como la ausencia de correlación entre los índices $h_{GA}$ y Scopus.

**Tabla 6.** Valores del análisis de correlación de los indicadores de tamaño, indexación y citación para las revistas del área general ciencias sociales. $r_s$ equivale a coeficientes de correlación de Spearman. Ns significa no significativo ($p > 0.05$).

| Indicador | $AIR_{IBNP}$ (Log$_{10}$) | $AIR_{GA}$ (Log$_{10}$) | $PI_{IBNP}$ | $PI_{LD}$ | $CR_{GA}$ (Log$_{10}$) | $CA\dot{x}_{GA}$ | $h_{GA}$ | $h_{SC}$ |
|---|---|---|---|---|---|---|---|---|
| $AIR_{IBNP}$ (Log$_{10}$) | 1.00 | 0.37 | 0.28 | 0.32 | 0.30 | Ns | Ns | Ns |
| $AIR_{GA}$ (Log$_{10}$) | 0.37 | 1.00 | Ns | 0.36 | 0.60 | 0.51 | 0.53 | Ns |
| $PI_{IBNP}$ | Ns | Ns | 1.00 | 0.73 | 0.46 | 0.37 | 0.41 | Ns |
| $PI_{LD}$ | 0.32 | 0.36 | 0.73 | 1.00 | 0.38 | Ns | 0.30 | Ns |
| $CR_{GA}$ (Log$_{10}$) | Ns | 0.60 | 0.46 | 0.38 | 1.00 | 0.93 | 0.92 | Ns |
| $CA\dot{x}_{GA}$ | Ns | 0.51 | 0.37 | Ns | 0.93 | 1.00 | 0.91 | Ns |
| $h_{GA}$ | Ns | 0.53 | 0.41 | 0.30 | 0.92 | 0.91 | 1.00 | Ns |
| $h_{SC}$ | Ns | Ns | Ns | Ns | Ns | Ns | Ns | 1.00 |

La prueba de independencia lineal de los indicadores de citación, (se excluyó el índice $h_{SC}$), mostró que un solo factor explicó la combinación lineal de las tres variables. De forma similar que en ciencias, el índice $h_{GA}$ explicó la mayor cantidad de varianza del factor (88.82 %) y su factor de carga o contribución única de cada variable al factor, fue igualmente el más alto (0.91).

Por otra parte, el análisis de regresión mostró resultados similares tanto en ciencias como en ciencias sociales, ya que la cantidad de citas por revista en GA ($y_1$) y el el índice $h_{GA}$ ($y_2$) dependieron linealmente de la cantidad de artículos en GA ($x_1$) y del puntaje de indexación de librería digital ($x_2$). La cantidad de citas por artículo no fue explicada por alguna de las variables consideradas.

En el área de ciencias, la regresión lineal múltiple de la cantidad de citas por revista en GA ($y_1$) ($R^2$ ajustado = 0,65; $F(2,86)=57.84$; $p < 0,001$) se explicó a través de la ecuación (Log$_{10}$ $y_1$= 0.133 + 0.717(Log$_{10}$ $x_1$) + 0.204 $x_2$). Así mismo, el índice $h_{GA}$ ($R^2$ ajustado = 0.54; $F(2,87) = 53.72$; $p < 0,001$) fue predicho por la ecuación: ($y_2$ = 0.086 + 0.539 (Log10 x1) + 0.361 x2).

En cuanto al área de ciencias sociales, la regresión múltiple de la cantidad de citas por revista en GA ($y_1$) ($R^2$ ajustado = 0,40; $F(2,74)=26.41$; $p < 0,001$) se predijo a través de la ecuación (Log$_{10}$ $y_1$ = 0.241 + 0.528 (Log$_{10}$ $x_1$) + 0.248 $x_2$). Por su parte, el índice $h_{GA}$ ($y_2$) fue predicho también por una ecuación similar ($R^2$ ajustado = 0.34; $F(2,80)=20.91$; $p < 0,001$), a través de la ecuación ($y_2$ = 0.126 + 0.46 (Log$_{10}$ $x_1$) + 0.254 $x_2$).

Esto puede interpretarse como: una mayor citación por revista depende de la combinación de hacer visibles muchos artículos en GA, y de la indexación en una mayor cantidad de Librerías digitales, o en unas pocas (solo Scopus)

**Clasificación de las revistas colombianas**

Las revistas de las áreas generales de ciencias y ciencias sociales, se organizaron en orden descendente con base en su valor de índice $h_{GA}$ (Tablas 7 y 8). También se incluyó la categoría de revistas IBNP y los valores del índice de citación promedio normalizado.

En las áreas de ciencias y ciencias sociales los cuartiles uno y dos (mediana) del índice $h_{GA}$ mostraron un valor igual, los cuales fueron: primer cuartil (h > 3), segundo cuartil (2 < h < 3), tercer cuartil (1 < h < 2) y cuarto cuartil (0 < h < 1). A continuación se presentan las revistas con valores de índice *h* del primer y segundo cuartil (h > 2).

**Tabla 7.** Clasificación de las revistas del área general de ciencias de acuerdo con su valor de índice *h* GA-PoP para el periodo 2003-2007.

| Clasificación | Título de la revista | $h_{GA}$ | Categoria IBNP | Citas promedio por articulo normalizado |
|---|---|---|---|---|
| 1 | COLOMBIA MÉDICA | 10 | A2 | 10.35 |
| 2 | LIVESTOCK RESEARCH FOR RURAL DEVELOPMENT | 8 | B | 5.28 |
| 3 | BIOMÉDICA | 7 | A1 | 3.48 |
| 4 | CALDASIA | 6 | A2 | 6.63 |
| 5 | INFECTIO | 6 | A2 | 8.41 |
| 6 | MEDUNAB | 6 | C | 3.38 |
| 7 | REVISTA DE SALUD PÚBLICA | 5 | A1 | 2.88 |
| 8 | REVISTA COLOMBIANA DE ENTOMOLOGÍA | 4 | A1 | 1.57 |
| 9 | AGRONOMÍA COLOMBIANA | 4 | A2 | 1.66 |
| 10 | AQUICHAN | 4 | A2 | 3.98 |
| 11 | IATREIA | 4 | A2 | 1.41 |
| 12 | REVISTA COLOMBIANA DE OBSTETRICIA Y GINECOLOGÍA | 4 | A2 | 0.80 |
| 13 | AVANCES EN ENFERMERÍA | 4 | C | 1.99 |
| 14 | BOLETÍN DE INVESTIGACIONES MARINAS Y COSTERAS | 3 | A2 | 0.59 |
| 15 | DYNA | 3 | A2 | 0.44 |
| 16 | INGENIERÍA E INVESTIGACIÓN | 3 | A2 | 0.95 |
| 17 | INVESTIGACIÓN Y EDUCACIÓN EN ENFERMERÍA | 3 | A2 | 0.90 |
| 18 | REVISTA COLOMBIANA DE CARDIOLOGÍA | 3 | A2 | 1.67 |
| 19 | REVISTA COLOMBIANA DE ESTADÍSTICA | 3 | A2 | 1.66 |
| 20 | REVISTA COLOMBIANA DE QUÍMICA | 3 | A2 | 2.36 |
| 21 | REVISTA GERENCIA Y POLÍTICAS DE SALUD | 3 | A2 | 1.82 |
| 22 | SALUD UNINORTE | 3 | A2 | 3.29 |
| 23 | VITAE | 3 | A2 | 0.91 |
| 24 | EARTH SCIENCES RESEARCH JOURNAL | 3 | B | 1.47 |
| 25 | INGENIERÍA Y UNIVERSIDAD | 3 | B | 3.83 |
| 26 | REVISTA EIA | 3 | B | 0.91 |
| 27 | REVISTA COLOMBIANA DE BIOTECNOLOGÍA | 3 | C | 0.82 |

**Tabla 8.** Clasificación de las revistas del área general de ciencias sociales de acuerdo con su valor de índice $h$ GA-PoP para el periodo 2003-2007.

| Clasificación | Título de la revista | $h_{GA}$ | Categoria IBNP | Citas promedio por articulo normalizado |
|---|---|---|---|---|
| 1 | REVISTA DE ECONOMÍA INSTITUCIONAL | 8 | A2 | 3.99 |
| 2 | UNIVERSITAS PSYCHOLOGICA | 7 | A1 | 3.78 |
| 3 | REVISTA LATINOAMERICANA DE PSICOLOGÍA | 5 | A1 | 3.89 |
| 4 | REVISTA COLOMBIANA DE PSICOLOGÍA | 5 | C | 3.79 |
| 5 | CUADERNOS DE DESARROLLO RURAL | 4 | A2 | 4.21 |
| 6 | DESARROLLO Y SOCIEDAD | 4 | A2 | 4.32 |
| 7 | ENSAYOS SOBRE POLÍTICA ECONÓMICA | 4 | A2 | 2.58 |
| 8 | REVISTA DE ESTUDIOS SOCIALES | 4 | A2 | 1.53 |
| 9 | LECTURAS DE ECONOMÍA | 4 | B | 3.37 |
| 10 | TABULA RASA | 4 | B | 2.15 |
| 11 | NÓMADAS | 4 | C | 1.32 |
| 12 | REVISTA COLOMBIANA DE SOCIOLOGÍA | 4 | C | 0.97 |
| 13 | ACTA COLOMBIANA DE PSICOLOGÍA | 3 | A1 | 1.95 |
| 14 | HISTORIA CRÍTICA | 3 | A1 | 1.43 |
| 15 | REVISTA COLOMBIANA DE PSIQUIATRÍA | 3 | A1 | 0.89 |
| 16 | ANÁLISIS POLÍTICO | 3 | A2 | 2.34 |
| 17 | AVANCES EN PSICOLOGÍA LATINOAMERICANA | 3 | A2 | 1.70 |
| 18 | CUADERNOS DE ADMINISTRACIÓN | 3 | A2 | 1.66 |
| 19 | ESTUDIOS GERENCIALES | 3 | A2 | 0.99 |
| 20 | INNOVAR | 3 | A2 | 0.74 |
| 21 | REVISTA COLOMBIANA DE EDUCACIÓN | 3 | B | 1.17 |
| 22 | REVISTA LASALLISTA DE INVESTIGACIÓN | 3 | B | 1.15 |
| 23 | SIGNO Y PENSAMIENTO | 3 | B | 1.29 |
| 24 | COYUNTURA ECONÓMICA | 3 | C | 1.97 |
| 25 | DESARROLLO Y SOCIEDAD | 3 | C | 1.46 |
| 26 | PALABRA CLAVE | 3 | C | 1.77 |
| 27 | PSICOLOGÍA DESDE EL CARIBE | 3 | C | 2.02 |
| 28 | REFLEXIÓN POLÍTICA | 3 | C | 0.92 |
| 29 | REVISTA COLOMBIANA DE ANTROPOLOGÍA | 3 | C | 1.51 |

**DISCUSIÓN**

**Indicadores de tamaño, indexación, citación y su interacción**

Se estimó la producción total del IBNP en ≈ 25 mil ítems, los cuales fueron indexados (visibles) en Google Académico ≈ 9 mil, recibiendo en total ≈ 3.5 mil citas. Asumiendo esta, como la producción doméstica (con alta proporción de autores colombianos) y contrastándola contra la producción internacional de autores con afiliación a instituciones colombianas en la Web of Knowledge (todas las áreas de conocimiento, tipos de documento y el mismo periodo de tiempo), donde se indexaron ≈ 5.5 mil ítems, recibiendo ≈ 15.7 mil citas, en más de 500 revistas; es posible afirmar que los artículos de investigadores colombianos en revistas internacionales, reciben mayor cantidad de citas que los artículos de revistas nacionales. Es decir, como un todo los artículos de investigadores colombianos en revistas internacionales poseen ≈ siete veces más citas que los artículos de revistas domésticas.

Este mismo patrón parece ser un fenómeno global y coincide con lo registrado en Latinoamérica (Figueira *et al.* 2003), China (Zhou & Leydesdorff 2007) y países de Europa oriental (Robu *et al.* 2001), donde la mayor parte de los resultados de investigación son publicados en revistas domésticas por razones como la pertinencia del conocimiento al contexto local (Gibbs 1995), barrera lingüística de los autores (Zhou & Leydesdorff 2007) o calidad intrínseca de los artículos (Arunachalam & Manorama 1989), teniendo en general un menor impacto que los artículos internacionales.

Por otra parte, en las áreas de ciencias y ciencias sociales el tamaño o productividad se asoció con variables de visibilidad (correlaciones débiles e intermedias), indicando que a mayor producción total (cantidad de artículos en IBNP), mayor producción visible (cantidad de artículos en GA). Esto refleja que GA no recuperó el contenido completo de cada una de las revista en la red, en este caso, recuperó el contenido total de cuatro revistas de ciencias y seis de ciencias sociales. Así mismo, este incremento en la producción visible se correlacionó con una mayor citación ($Log_{10}$ cantidad de citas por revista GA e índice $h_{GA}$). Las razones por las cuales GA recupera o no el contenido de una revista o ítem en la red, excede los objetivos de este trabajo; sin embargo, en este aspecto, GA ha recibido críticas en torno a la calidad de sus datos (Jacso 2005).

En cuanto a la inclusión de revistas colombianas en librerías digitales, GA indexó la mayor cantidad de revistas y artículos. En el área de ciencias, GA y Scielo indexaron una cantidad mayor de revistas que las restantes. En ciencias sociales, Redalyc indexó una mayor cantidad de revistas que Google Académico y restantes librerías digitales. Estas tendencias eran las esperadas, ya que Scielo hace énfasis en revistas de ciencias biomédicas (Packer *et al.* 2001) y Redalyc lo hace en ciencias sociales (Aguado López *et al.*).

Las revistas colombianas de ciencias y ciencias sociales indexadas en Scopus, presentaron una mayor citación promedio que las revistas indexadas en las restantes librerías digitales. Aunque GA indexó ≈ seis veces más artículos que Scopus, una revista indexada en Scopus recibió en promedio ≈ dos veces más citas que una revista indexada en Google Académico. El promedio de citas por artículo, sin embargo, no fue diferente entre librerías digitales en el área de ciencias, pero si en ciencias sociales donde fue mayor para Scopus.

Respecto a las categorías de las revistas del IBNP, la producción total de las revistas categoría C fue mayor que en las restantes categorías; sin embargo, cada revista de categoría A1-A2 produjo en promedio más artículos que una revista categoría B-C. Así mismo, las revistas categorías A1-A2 presentaron indicadores mayores de citación (citas promedio por revista y por artículo) e indexación (puntaje de indexación de librería digital) que las revistas B-C. Estos resultados implican que las actuales categorías (A1, A2, B o C) del IBNP, o la pertenencia a una mayor cantidad de librerías digitales, no reflejan una mayor citación o un mayor uso de los artículos de una revista por parte de la comunidad.

En términos de nuestro trabajo, las variables que explicaron la citación se aclararon en el análisis de regresión, mostrando que a mayor puntaje de indexación de librería digital y mayor su producción visible, mayores los indicadores de citación (cantidad de citas por revista GA e índice $h_{GA}$). Esto refleja que el incremento (no lineal) en la cantidad de citas por revista ocurre por un efecto combinado entre la visibilidad en GA y la pertenencia a ciertas librerías digitales (particularmente Scopus). Aspecto acorde a lo planteado por Leydesdorff (2009), quien describió que la influencia de una revista es la combinación entre impacto y productividad.

La estructura de la función Log-lineal establecida, es acorde con lo planteado en la teoría de ciencias de la información como leyes de distribución de potencia (Milojevic 2010; Newman 2005; van Raan 2006b). Este patrón no lineal es semejante a lo presentado por Katz (1999), quien demostró para un conjunto de datos en 152 sub-áreas de conocimiento, que existió una relación de dependencia no lineal entre el tamaño de una revista y su citación, constante e independiente del tamaño o nacionalidad. Así mismo, Katz (1999) planteó que este tipo de distribución, describe situaciones donde existen ventajas acumulativas o "efecto Matthew", en el cual aquellos sujetos con gran presencia en la comunidad, ganan mayor reconocimiento comparados con aquellos con poco reconocimiento. Estos sujetos pueden ser grupos de investigación (van Raan 2006b), instituciones, revistas o países (Katz 1999). En términos de revistas, en la medida que las revistas participen más en el sistema de ciencia, ganarán una ventaja acumulativa e incrementarán sus beneficios. Visto de otra forma, para una revista con poco tamaño o poco visible, la probabilidad de ser citada no depende en aumentar su producción, sino de otros factores. Aunque nuestro estudio predijo la citación promedio por revista y el índice $h_{GA}$, reconocemos que además de la visibilidad, las motivaciones que tiene un lector para citar un trabajo están asociadas con la calidad del mismo y otros factores (Egghe & Rousseau 1990; Leydesdorff 1998).

**Clasificación de las revistas colombianas**

Se estimaron para 173 revistas del IBNP indicadores de citación; sin embargo, debido a la ausencia de independencia lineal entre estos indicadores, se seleccionó el índice $h_{GA}$ como el mejor indicador, ya que explicó la mayor cantidad de varianza del factor y porque permite su comparación con cualquier revista en el ámbito latinoamericano o internacional.

El cálculo del índice $h$ plantea una relación entre el número de artículos y su citación (Hirsch 2005). En nuestro caso, el índice $h_{GA}$, fue predicho en algún grado por las citas promedio por revista y por su visibilidad (cantidad de artículos en GA). Ya que el índice $h$ no tiene en cuenta toda la producción de una revista (se excluyen los artículos no citados) (Norris & Oppenheim 2010), es nuestro caso, un buen indicador de visibilidad en GA porque excluye además, los artículos no indexados. No obstante, sugerimos la interpretación con precaución de este índice en el área de ciencias sociales, ya que no existió una correlación positiva entre los índices $h_{GA}$ y $h_{SC}$, como sí ocurrió en ciencias. Esta relación puede ser producto del poco número de revistas indexadas de ciencias sociales en Scopus para el momento del muestreo y futuros trabajos pueden dilucidar la ausencia o presencia de esta relación.

Sin embargo, varios trabajos que han comparado indicadores de citación calculados en WoS, Scopus y GA, han mostrando resultados variables. Por ejemplo, Kulkarni et al (2009) encontraron que GA y Scopus recuperan mayor cantidad de citas que WoK en revistas de medicina general. En términos del índice $h$, Saad (2006) comparó entre WoK y GA este índice en autores de la revista Journal of Consumer Research durante 1989-2005 y encontró, como Harzing (2009) comparando 838 revistas de negocios y economía, una buena correlación entre los índices $h$ de las dos librería digital, con una tendencia a la sobreestimación en GA.

El índice $h$ posee varias ventajas: (i) facilidad de interpretación (una revista con un índice $h$ ha publicado $h$ artículos, cada uno citado al menos $h$ veces); (ii) se actualiza casi en tiempo real a través de GA; (iii) identifica aquellos sujetos (revistas) que consistentemente producen un flujo de buenos trabajos sostenidamente en el tiempo, sobre aquellos que producen muchos trabajos, pero poco citados (Hirsch 2005) y (iv) captura las dimensiones ortogonales tamaño e impacto en un solo indicador (Leydesdorff 2009).

Esta última característica, es a la vez una de sus desventajas, porque puede mal interpretar su valor (Leydesdorff 2009). Por ejemplo, una revista con *h* = 5 puede tener cinco artículos con cinco citas cada uno, sin importar cuantos artículos hayan sido publicados o escritos y citados por debajo de ese número, castigando la alta productividad. De forma similar, una revista con *h* = 5 puede tener cinco artículos altamente citados, sin importar el nivel de citación (Norris & Oppenheim 2010) premiando la baja productividad. Otra desventaja, implica que una vez, una revista ha adquirido un índice *h* este no decrece, aún si la revista está inactiva, con el peligro potencial de disminuir su calidad en el tiempo (Norris & Oppenheim 2010).

Bajo la perspectiva de Hirch (2005) se sugiere que dos sujetos (revistas) con índices *h* similares, son comparables en términos de su impacto científico, aún si son muy diferentes el número total de artículos o el número total de citas. Así mismo, Hirch (Hirsch 2007) sugirió que el índice *h* tiene un carácter predictivo, ya que potencialmente podría ayudar a identificar académicos (revistas) que serán exitosos en el futuro.

El método de estimar el índice a través de Publish or Perish tiene dos desventajas. La imposibilidad de remover la auto-citación, aspecto que puede generar variaciones en el índice (Norris & Oppenheim 2010) y el tiempo que consume la verificación extensiva y corroboración de cada una de las referencias individualmente, que hacen al proceso de análisis excesivamente largo, en comparación con Scopus o WoK (Jacsó 2008), por lo cual la consulta de información debe hacerse en un lapso corto de tiempo.

Así mismo, se presentó el índice de citación promedio normalizado por artículo, el cual es aplicable solo al contexto nacional, porque utilizó exclusivamente títulos colombianos y provee de información valiosa al editor sobre su posición respecto al promedio nacional. Los resultados del índice normalizado fueron acordes con lo planteado por Van Raan (2006a), quien mostró que los indicadores normalizados de 147 grupos de investigación en química se correlacionaron con el índice *h*. No obstante las bondades en términos de comparación, este indicador ha sido criticado, porque el promedio no es un estadístico adecuado para describir distribuciones sesgadas, típicas de los valores de citación (van Raan 2006b). Aunque el promedio normalizado hace que los artículos con alto impacto se promedien, este indicador se correlaciona adecuadamente con otros estadísticos como la mediana y el porcentaje de artículos no citados (Moed *et al.* 1995). Sin embargo, es preciso reconocer que indicadores normalizados deben ser analizados independientemente en cada área de conocimiento (Moed *et al.* 1995).

La nueva clasificación presentada es en sí misma controversial, porque sensu stricto no refleja el uso que la comunidad académica hace de las revistas. Aunque existió una asociación positiva entre las categorías IBNP y el índice h, la clasificación reacomoda revistas categorías A2, B y C en los primeros lugares. En ciencias por ejemplo, en los cinco primeros lugares solo existe una revista A1 y el segundo lugar es una revista categoría B. En Ciencias sociales el primer lugar es una revista A2 y el cuarto una revista categoría C. El índice *h*, se ha reportado puede rescatar de la oscuridad a aquellos investigadores (revistas) que han hecho contribuciones significativas al conocimiento pero que no han ganado suficiente reputación para reconocerlo (Van Raan 2006a).

**Implicaciones en políticas públicas**

Un artículo científico aparece en una revista académica después de atravesar por un proceso de revisión por pares; el aura de autoridad adquirida, que implica ser publicado en revistas con prestigio, refleja un grado de dificultad que conduce a ganar reconocimiento por la comunidad (Darch & Underwood 2005; Van Leeuwen et al. 2003).

Colombia se encuentra en una etapa de transición desde un sistema de divulgación e incentivos que hace énfasis en las publicaciones nacionales (Charum 2004) (cerrado y endogámico), hacia uno visible internacionalmente que se soporta en redes de conocimiento y medido por indicadores de impacto.

Instituciones e investigadores colombianos están en camino a convertir en objetivo institucional la visibilidad internacional de sus resultados de investigación, por medio de su divulgación en revistas indexadas particularmente Scopus, la cual actualmente indexa una mayor cantidad de revistas colombianas en comparación con WoK. Aunque esta decisión es controversial, puede ser positiva en varios aspectos. En primer lugar, porque las instituciones mejorarían su posición en la clasificación mundial de universidades y pueden evaluar rápidamente el desempeño de sus docentes e investigadores. Por último, para investigadores, docentes y estudiantes se esperaría un efecto positivo sobre la calidad de la docencia, investigación y posibilidades de financiación.

Esta nueva estrategia de visibilidad del conocimiento científico colombiano (e.g. http://tinyurl.com/2e7wzek), puede tener importantes repercusiones sobre el actual IBNP. Toca aspectos clave, como los mecanismos que garantizan el uso correcto de los recursos públicos destinados a investigación. Por ejemplo, el costo por suscripciones a grandes multinacionales del conocimiento vs costo de perder recursos públicos en investigaciones de baja calidad e invisible, según Gibbs (1995) el fenómeno de ciencia pérdida.

El IBNP fue creado para controlar la calidad de las revistas colombianas y para dar soporte al incremento salarial de profesores e investigadores del estado. El IBNP clasifica las revistas midiendo criterios de calidad científica, calidad editorial, estabilidad, visibilidad y reconocimiento nacional e internacional, en términos de características editoriales internacionalmente adoptadas.

El conocimiento científico colombiano es en sí mismo valioso, no obstante, las revistas que no inicien estrategias y acciones para mejorar su visibilidad e indicadores, están amenazadas. En tal sentido, sugerimos estas deben estar encaminadas a aumentar el uso del conocimiento científico por la toda comunidad, incluyendo el estado, ya que el conocimiento de las revistas colombianas no es empleado consistentemente para tomar decisiones de gobierno. Sin embargo, es muy importante reconocer que el modelo basado en indicadores (citacionista) puede no aplicar a la realidad colombiana (¿un artículo muy citado resuelve problemas de violencia, corrupción y pobreza?). Así mismo, es pertinente reconocer que tanto las publicaciones colombianas como investigadores se encuentran en desventaja (León-Sarmiento *et al.* 2005) respecto a sus pares internacionales, particularmente por la poca inversión destinada del PIB a investigación.

Aunque numerosas revistas han solicitado la inclusión en librerías digitales como Redalyc, Scielo o Latindex, proponemos otras estrategias como un portal colombiano de conocimiento, donde se integren revistas (e.g http://www.koreamed.org), políticas públicas, problemas sociales e hipótesis, a través del cual, empleando herramientas de análisis basado en redes semánticas (e.g. http://hypothesis.alzforum.org/) se formulen explícitamente y propongan soluciones a los problemas críticos de Colombia, cumpliendo de tal forma con la Ley de Ciencia y Tecnología.

Así mismo, notamos que las revistas colombianas carecen de mecanismos eficientes de divulgación relacionados con tecnologías de la Red 2.0, tales como URI (Uniform Resource

Identifiers) o DOI (Digital Object Identifiers), así como sistemas de envío electrónico de manuscritos y revisión por pares (e.g. Open Journal System), sistemas automáticos de detección de plagio y la puesta en práctica iniciativas como Open ID o Research ID para evitar homonimia. Estas tecnologías llevan a un aumento de la visibilidad porque facilitan la citación y descarga automática de archivos PDF a través de software como EndNote, Mendely, Jabref o Mekentosj Papers (Hull et al. 2008; Renear & Palmer 2009).

Finalmente, en los próximos años podrá establecerse si la estrategia de internacionalización ("publique o perezca, en idioma inglés, en una revista Scopus, en colaboración internacional y resolviendo problemas críticos de la sociedad colombiana") tendrá un efecto positivo sobre la calidad de la educación, investigación e innovación, dando a la par soluciones a los problemas críticos de Colombia. Ya que el desarrollo de un país depende en gran medida de su fortaleza científica y en la habilidad de resolver problemas en áreas como salud pública, enfermedades infecciones, manejo ambiental o progreso industrial (Kirsop & Chan 2005), otros indicadores como el uso de las publicaciones académicas en la formulación de políticas públicas, leyes, manejo de recursos naturales y control de recursos públicos, puede ser más importante que el factor de impacto *per se*.

**CONCLUSIONES**

Los métodos empleados permitieron realizar una clasificación basada en el índice *h* para las revistas colombianas de las áreas generales de ciencias y ciencias sociales, no incluidas en Web of Science o Scopus. Se identificaron algunos patrones relevantes para comprender la dinámica de citación de las revistas colombianas.

- Las revistas mejor clasificadas IBNP son más productivas, pero no necesariamente las más citadas.
- Las revistas indexadas en Scopus tienen más citas que aquellas indexadas exclusivamente en Redalyc, Scielo o Google Académico.
- Las revistas más visibles en Google Académico son más citadas.
- La producción total, no está asociada con la citación.
- La actual clasificación del IBNP, cambia al aplicar indicadores de citación.

- Las revistas académicas deben incorporar estrategias y acciones para mejorar su visibilidad, las cuales pueden ser monitoreadas en el tiempo con indicadores de citación